\documentclass[
reprint,
 aps,
 superscriptaddress,
bibnotes,
 prl,
floatfix]{revtex4-2}

\usepackage{physics}
\usepackage{dsfont}
\usepackage{color,newfloat}
\usepackage{graphicx}
\usepackage{dcolumn}
\usepackage{bm}
\usepackage{amsmath,amssymb,amsthm}
\usepackage{mathtools}
\usepackage{multirow}
\usepackage{hhline}
\usepackage{comment}
\usepackage[dvipsnames]{xcolor}
\usepackage{hyperref}
\usepackage{wasysym}
\hypersetup{
    colorlinks=true,
    linkcolor=Maroon,
    citecolor=Maroon,
    urlcolor=Maroon
}

\usepackage[caption = false]{subfig}
\usepackage{lineno}
\usepackage{siunitx}

\usepackage{makecell}

\begin{document}


\title{
Photonic fusion of entangled resource states from a quantum emitter
}


\author{Yijian Meng}
\affiliation{Center for Hybrid Quantum Networks (Hy-Q), Niels Bohr Institute, University of Copenhagen, Blegdamsvej 17, Copenhagen 2100, Denmark}

\author{Carlos F.D. Faurby}
\affiliation{Center for Hybrid Quantum Networks (Hy-Q), Niels Bohr Institute, University of Copenhagen, Blegdamsvej 17, Copenhagen 2100, Denmark}

\author{Ming Lai Chan}
\affiliation{Center for Hybrid Quantum Networks (Hy-Q), Niels Bohr Institute, University of Copenhagen, Blegdamsvej 17, Copenhagen 2100, Denmark}

\author{Patrik I. Sund}
\affiliation{Center for Hybrid Quantum Networks (Hy-Q), Niels Bohr Institute, University of Copenhagen, Blegdamsvej 17, Copenhagen 2100, Denmark}

\author{Zhe Liu}
\affiliation{Center for Hybrid Quantum Networks (Hy-Q), Niels Bohr Institute, University of Copenhagen, Blegdamsvej 17, Copenhagen 2100, Denmark}

\author{Ying Wang}
\affiliation{Center for Hybrid Quantum Networks (Hy-Q), Niels Bohr Institute, University of Copenhagen, Blegdamsvej 17, Copenhagen 2100, Denmark}

\author{Nikolai Bart}
\affiliation{Lehrstuhl f{\"u}r Angewandte Festk{\"o}rperphysik, Ruhr-Universit{\"a}t Bochum, Universit{\"a}tsstrasse 150, D-44780 Bochum, Germany}

\author{Andreas D. Wieck}
\affiliation{Lehrstuhl f{\"u}r Angewandte Festk{\"o}rperphysik, Ruhr-Universit{\"a}t Bochum, Universit{\"a}tsstrasse 150, D-44780 Bochum, Germany}

\author{Arne Ludwig}
\affiliation{Lehrstuhl f{\"u}r Angewandte Festk{\"o}rperphysik, Ruhr-Universit{\"a}t Bochum, Universit{\"a}tsstrasse 150, D-44780 Bochum, Germany}

\author{Leonardo Midolo}
\affiliation{Center for Hybrid Quantum Networks (Hy-Q), Niels Bohr Institute, University of Copenhagen, Blegdamsvej 17, Copenhagen 2100, Denmark}

\author{Anders S. Sørensen}
\affiliation{Center for Hybrid Quantum Networks (Hy-Q), Niels Bohr Institute, University of Copenhagen, Blegdamsvej 17, Copenhagen 2100, Denmark}

\author{Stefano Paesani}
\email{stefano.paesani@nbi.ku.dk}
\affiliation{Center for Hybrid Quantum Networks (Hy-Q), Niels Bohr Institute, University of Copenhagen, Blegdamsvej 17, Copenhagen 2100, Denmark}
\affiliation{NNF Quantum Computing Programme, Niels Bohr Institute, University of Copenhagen, Blegdamsvej 17, Copenhagen 2100, Denmark.}

\author{Peter Lodahl}
\email{lodahl@nbi.ku.dk}
\affiliation{Center for Hybrid Quantum Networks (Hy-Q), Niels Bohr Institute, University of Copenhagen, Blegdamsvej 17, Copenhagen 2100, Denmark}

\date{}


\begin{abstract}
Fusion-based photonic quantum computing architectures rely on two primitives: i)  near-deterministic generation and control of constant-size entangled states and ii) probabilistic entangling measurements (photonic fusion gates) between entangled states. 
Here, we demonstrate these key functionalities by fusing resource states deterministically generated using a solid-state spin-photon interface.
Repetitive operation of the source leads to sequential entanglement generation, whereby curiously entanglement is created between the quantum states of the same spin at two different instances in time. 
Such temporal multiplexing of photonic entanglement provides a resource-efficient route to scaling many-body entangled systems with photons.

\end{abstract}

\date{\today}

\maketitle

\begin{figure*}
    \centering
    \includegraphics[width=1.0\textwidth]{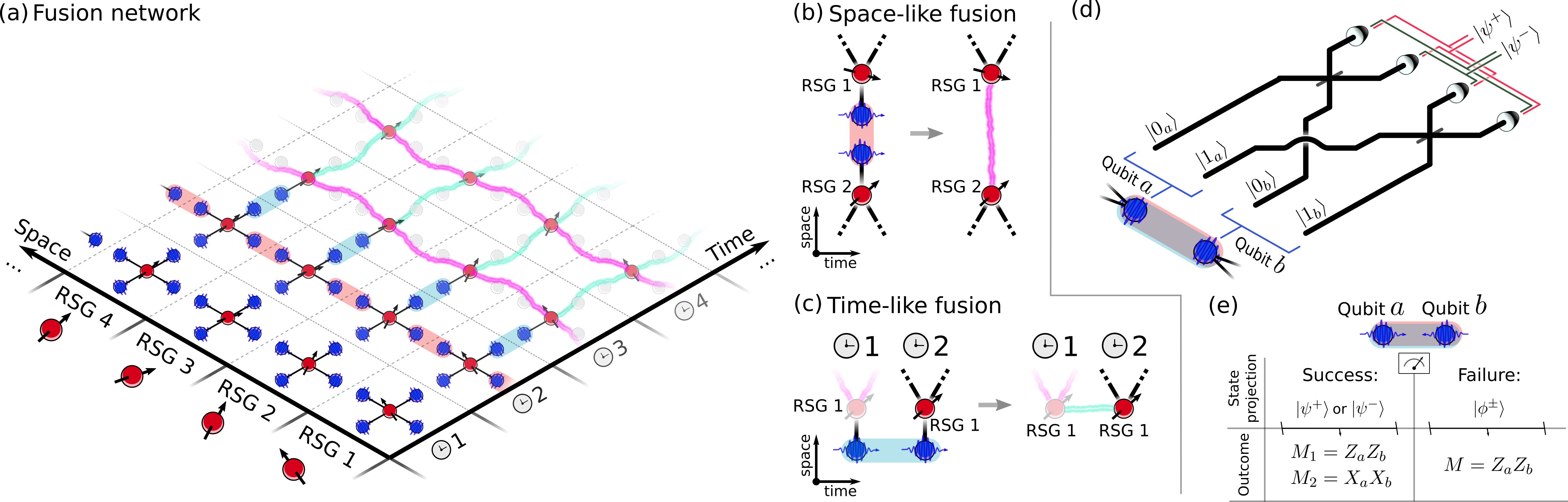}
    \caption{
   \textbf{Fusion-based quantum photonic systems.}
    (a) Example of a fusion network generating long-range quantum correlations by sequentially fusing photons from different resource states. (\clock 1) Resource state generators (RSG), here depicted as quantum emitters, are used to generate constant-size resource states of entangled photons. (\clock 2) Space-like fusions (red shaded) are used to fuse photons emitted in the same clock cycle  (b), and (\clock 3) time-like fusions (blue shaded) fuse photons emitted from the same RSG but at different times using a temporal delay on the earliest photon (c). (\clock 4) The fusion network results in the generation of space-like (purple) and time-like (cyan) quantum correlations.
    (d) Schematic of a fusion measurement implemented via a probabilistic linear-optical Bell state analyzer with a success probability of 50\%. The four two-photon detection patterns corresponding to successful fusion outcomes ($\psi^\pm$ projections) are shown, while the remaining two-photon patterns are associated with fusion failure ($\phi^\pm$ projections).
    (e) Projective states and parity outcomes for a fusion measurement implemented via probabilistic linear-optical Bell state analyzer for the fusion success and failure cases, with $\ket{\psi^\pm} = (\ket{01} \pm \ket{10})/\sqrt{2}$ and $\ket{\phi^\pm} = (\ket{00} \pm \ket{11})/\sqrt{2}$.
    }
    \label{fig:1}
\end{figure*}

\noindent\textbf{Introduction.}
Quantum computing relies on the realization of a universal set of one- and two-qubit gate operations.
In photonics, the lack of photon-photon interactions makes deterministic two-qubit gates challenging.
This limitation has motivated the development of alternative quantum computing approaches tailored to the photonic platform where entangling gates can be probabilistically implemented through measurements~\cite{Raussendorf2006,Browne2005,GimenoSegovia2015}.  
In this context, fusion-based quantum computing (FBQC) has emerged as a new and resource-efficient approach ~\cite{Bartolucci2023} where photons are continuously created in small entangled resource states and rapidly measured in shallow linear-optics circuits. 
Nonetheless, the quantum information survives in the system via quantum teleportation through \textit{fusion gates} --- i.e. entangling two-photon measurements that may be implemented probabilistically with linear-optical circuits~\cite{Browne2005}.
In FBQC, the quantum computing backbone is a fusion network consisting of multiple entangled resource states that are routed from the sources and fused together.  
 Fig.~\ref{fig:1}a illustrates an example of a fusion network of spin-photon entangled resource states \cite{cogan_deterministic_2023,coste_high-rate_2023,thomas_efficient_2022,meng2023deterministic}. 
The fusion operations either proceed in \textit{space} where two separate resource states are combined (see Fig.~\ref{fig:1}b) or in \textit{time} where photons from the same source but emitted at different times are fused (see Fig.~\ref{fig:1}c).
 The latter approach applies an optical delay (e.g., in an optical fiber) to interleave subsequently emitted photons, which may offer a significant resource reduction in the required number of physical photon sources~\cite{bombin2021interleaving}. 
Significant progress has been reported on developing FBQC photonic architectures tailored to hardware capabilities and the physical noises and operations~\cite{Bell2023, Sahay2023, Bombin2023, Paesani2023, bombin2023fault}.

A central challenge in FBQC is generating the required initial entangled resource states. 
All-optical approaches use photonic circuits and multiplexing to convert heralded probabilistic linear-optical processes into near-deterministic entanglement generation. 
However, they require immense hardware overheads that render them highly challenging for near-term technologies~\cite{Migdall2002, bartolucci2021creation}.
Quantum emitters emerge as a platform with a strong potential to surpass these limitations by naturally enabling the deterministic generation of photonic entanglement~\cite{Lindner2009, Economou2010}. 
Investigated quantum emitter platforms include quantum dots (QDs)~\cite{Schwartz2016, Istrati2020, Appel2022,coste_high-rate_2023, meng2023deterministic}, atoms~\cite{Yang2022, thomas_efficient_2022}, and color centers~\cite{vasconcelos2020}, and rely on creating an efficient spin-photon interface where spin-dependent photon generation deterministically entangles the emitted photonic qubits~\cite{Lindner2009}.
Recently, photonic resource states with up to 14 photonic qubits and high-fidelity was demonstrated with a quantum emitter~\cite{thomas_efficient_2022}.
%
%
However, combining deterministic resource state generation with photonic circuitry implementing FBQC fusion operations has so far been elusive for photonic quantum technologies.

We demonstrate the key stepping stones of FBQC, i.e. deterministic photonic resource state generation and photonic fusion with a solid-state platform based on QD sources.
QDs are considered a particularly promising platform for this approach since they are capable of generating long strings of highly indistinguishable photonic qubits at a GHz repetition rate~\cite{uppu2020scalable, tomm2021}, allow high indistinguishability between QDs \cite{zhai2022,papon2023independent}, and are compatible with integrated photonic circuits for processing quantum information \cite{Sund2023, Wang2023, maring2023general}.
The present work realizes temporal fusion operations, i.e. we fuse consecutive resource states generated from the same QD but at different times. 
In this process, the state of a single electron spin in the QD becomes entangled with itself, but at two different instances in time. 
This peculiar entanglement phenomenon is an asset in FBQC since it reduces the required overhead on the number of matter qubits in the architecture~\cite{bombin2021interleaving}. 

\ \\

\begin{figure}
    \centering
    \includegraphics[width=0.45\textwidth]{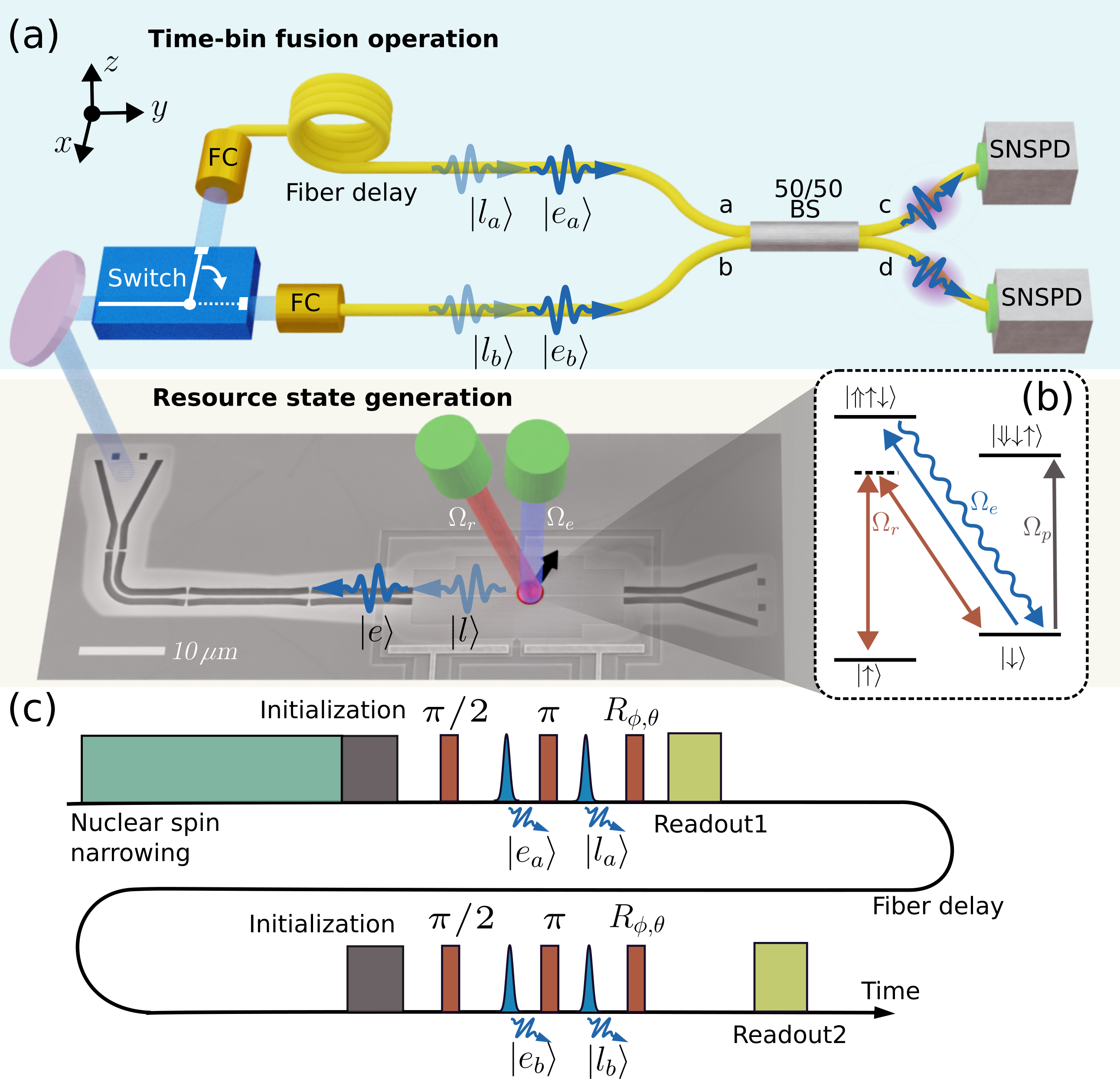}
   \caption{\textbf{Experimental setup.} (a) Schematic diagram of the experiment. The photons generated at time $a$ and $b$, 300~ns apart, are entangled with the same quantum dot spin embedded in a photonic crystal waveguide.   An electro-optic modulator switches the first photon into path $a$ and the second to path $b$. In path $a$, a fiber coupler (FC) collects the photon into a \SI{300}{ns} fiber delay.  The two photons overlap in time at a 50/50 beamsplitter. The joint detection of two photons in path $c$ or $d$ heralds an entangled state of the spin.  (b) Energy level diagram of the quantum dot spin with an electron groundstate spin of $\ket{\uparrow}$ or $\ket{\downarrow}$. Controlled Rabi oscillations of the spin state can be achieved through a Raman laser $\Omega_r$. (c) Pulse sequence applied to the quantum dot spin. Before initialization of the spin, nuclear spin narrowing is performed to increase the spin-coherence time. The spin-photon entanglement consists of a $\pi/2$ and a $\pi$ rotation together with the emission of a photon in different time bins. The spin state can be measured in different bases by rotating the spin before readout. } 
    \label{fig:2}
\end{figure}

\noindent\textbf{Experimental scheme.}
The experimental setup is outlined in Fig. \ref{fig:2}a.
It consists of a resource state generator chip, an active switch for routing photons from the emitted resource states, and a fiber-based temporal fusion gate via fiber delay and interference. 
The resource state generator is implemented using a spin-photon interface in an InGaAs QD that is embedded in a GaAs photonic crystal waveguide (PCW) (see electron microscope image in Fig.~\ref{fig:2}a). 
The QD possesses an optically cycling transition at \SI{947.86}{nm} and is driven resonantly ($\Omega_e$) for deterministic generation of single photons (blue excitation in Fig.~\ref{fig:2}) with a near-unity collection efficiency into the PCW~\cite{arcari2014near}.  
The QD is deterministically charged with a single electron spin through a bias voltage and we apply an external magnetic field of $\SI{4}{T}$ along $+y$ direction (Voigt geometry) to access the two Zeeman spin ground states $\ket{\downarrow}$ and $\ket{\uparrow}$, see Fig. \ref{fig:2}b. 
Coherent Rabi spin rotations are implemented by driving the QD with an off-resonant Raman laser $\Omega_r$ at \SI{650}{GHz} red-detuning from the optical transition.
In advance, the spin coherence time is increased by mitigating the nuclear spin noise bath by implementing optical cooling at the start of each experimental round~\cite{meng2023deterministic,gangloff_quantum_2019}. 

The protocol proceeds by operating spin-selective photon emission processes to deterministically generate entangled resource states by sequentially emitting photons~\cite{Lindner2009}.
In the present case, the photonic qubit is defined by whether the photon is emitted in an early ($\ket{e}$) or a late ($\ket{l}$) time bin, and a spin-echo reshaping $\pi$-pulse ensures that the protocol is robust towards spin dephasing~\cite{Tiurev2022}.
The pulse sequence is shown in Fig. \ref{fig:2}c and is repeated after 300 ns to generate two separate resource states for the fusion experiment.
Following the nuclear spin narrowing pulse, the spin state is initialized by resonantly exciting the optical transition with a laser pulse $(\Omega_p)$ followed by a sequence of alternating spin rotations and optical excitation/emission processes. 
Subsequently, the spin state is read out by \SI{200}{ns} of optical pumping of the diagonal transition $\ket{\downarrow}\rightarrow\ket{\Uparrow\uparrow\downarrow}$, see Fig. \ref{fig:2}b.
This procedure leads to photon emission if the spin is in the state $\ket{\downarrow}\equiv \ket{1}$, leading to measuring the spin state in the computational (Pauli $Z$) basis. 
To measure the spin state in other Pauli bases, we use a rotating pulse $R_{\phi,\theta}$ at the end of the resource state generation sequence and prior to the optical spin pumping. 
The spin is then reinitialized to generate a second resource state and is read out a second time.
The fusion experiment initially generates two separate spin-photon entangled resource states at different instances of time $t_a$ and $t_b$ ($t_b-t_a = 300$~ns),  i.e. $\ket{\psi^-}(t_i)=(\ket{0_i}\ket{e_i}-\ket{1_i}\ket{l_i})/\sqrt{2}$,  $i=a,b$. Here the qubit state $\ket{0}$ ($\ket{1}$) corresponds to the $\ket{\uparrow}$ ($\ket{\downarrow}$) spin state, and $\ket{e}$ ($\ket{l}$) denotes the emitted photon occupying the early (late) time-bin.
Subsequently, the first photon (labeled $a$) and the spin readout signal are routed with an electro-optic switch to a single-mode fiber delay matching the \SI{300}{ns} time difference between the two resource state generation times. 
The second photon is routed to a single-mode fiber without implementing a delay, resulting in both time-bin photonic qubits arriving simultaneously at a balanced fiber beam splitter (BS), as shown in Fig.~\ref{fig:2}a.
Photon detection on the output modes is implemented via superconducting-nanowire single-photon detectors (SNSPDs) that resolve the early and late photon arrival times.
In this circuit, the two early components $\ket{e_a}$ and $\ket{e_b}$ of the two time-bin qubits interfere at the BS, and the same is the case for the late components $\ket{l_a}$ and $\ket{l_b}$.
By identifying $\ket{e_i}$ ($\ket{l_i}$) with $\ket{0_i}$ ($\ket{1_i}$) as the computational state of each time-bin photonic qubit, the implemented scheme corresponds to a time-bin implementation of the photonic fusion circuit in Fig.~\ref{fig:1}e.
Labeling the two output modes of the BS as $c$ and $d$ (see Fig.~\ref{fig:2}a), the successful fusion outcomes correspond to measuring the two photons in the detection patterns $\ket{e_c l_d}$ and $\ket{e_d l_c}$ for a Bell state projection into $\psi^-$, and $\ket{e_c l_c}$ and $\ket{e_d l_d}$ for projecting into $\psi^+$.
The remaining detection patterns are associated with fusion failure, i.e. projection of the joint state of the two photons into the subspace spanned by the Bell states $\phi^\pm$.

\begin{figure}
    \centering
    \includegraphics[width=0.48\textwidth]{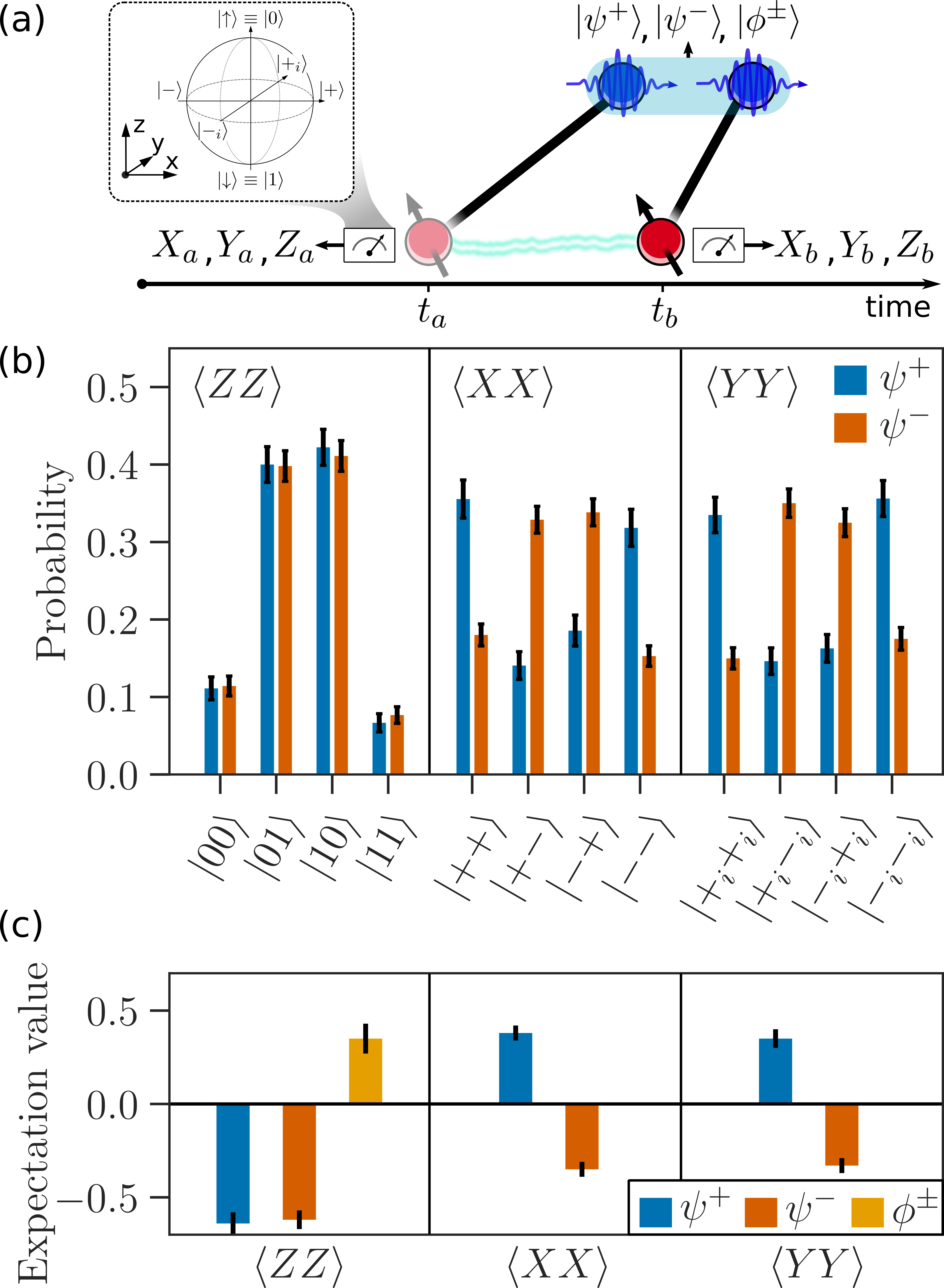}
    \caption{ \textbf{Fusion measurement results.} (a) The performance of the time-like fusion is probed by analyzing the entanglement fidelity of the two spin states after fusing the two photons in the resource states. (b) Spin entanglement correlation measurements are shown for the successful fusion outcomes where the photons are projected into $\psi^+$ (blue) and $\psi^-$ (red) for the different eigenstates of the spin Pauli basis: $\{\ket{0} \equiv \ket{\uparrow}, \ket{1} \equiv \ket{\downarrow}\}$ the eigenbasis of $Z$, $\{\ket{\pm}= (\ket{0} \pm \ket{1})/\sqrt{2}$\} the eigenbasis of $X$, and $\{\ket{\pm_i}= (\ket{0} \pm i\ket{1})/\sqrt{2}\}$ the eigenbasis of $Y$. (c) Expectation values of the spin state stabilizers conditioned on success ($\psi^\pm$) and failure ($\phi^\pm$) of the photonic fusion. Due to the erasure of the $XX$ and $YY$ fusion outcomes upon failure, only the $ZZ$ value is shown for this case. All error bars are estimated from Poissonian photon statistics.}
    \label{fig:3}
\end{figure}

\ \\

\ \\
\noindent\textbf{Results.}
A photonic fusion gate consumes two fused photons from different resource states to generate quantum correlations between the remaining qubits.
In our case, these correlations are generated between the two spin qubits encoded in the QD spin at times $t_a$ and $t_b$, as depicted in Fig.~\ref{fig:1}c, representing a time-like fusion operation.
The two resource states are spin-photon Bell states $\ket{\psi^-}(t_i)$ and a successful fusion measurement projects the photons into $\psi^+$ ($\psi^-$) to generate the joint two-spin states $\ket{\psi^\pm}_s = (\ket{\uparrow_a \downarrow_b} \pm \ket{\downarrow_a \uparrow_b})/\sqrt{2}$. 
This corresponds to a spin entangled with itself at two different instances of time.
These states are stabilized by the joint Pauli operators $-ZZ$, $\pm XX$, and $\pm YY$, i.e. each represents the unique common eigenstate of eigenvalue +1 for these commuting operators~\cite{gottesman1997stabilizer}.
Upon fusion failure ($\phi^\pm$ projection), the $ZZ$ fusion outcome is still obtained but the $XX$ and $YY$ outcomes are erased. 
This results in the generation of perfect correlations between the spin qubits only in the $ZZ$ basis (with expectation value $+1$), but no correlations in the $XX$ and $YY$ bases.   
The correspondence between the fusion outcomes and the resulting joint state of the spin qubits described above enables us to probe the fusion operation by measuring the quantum correlations generated between the spin qubits.
Such an analysis is performed by measuring the states of the two spin qubits in different single-qubit Pauli bases to obtain the shared quantum correlations, as depicted in Fig.~\ref{fig:3}a. 
The measured correlations between the spin qubits for the $ZZ$, $XX$, and $YY$ Pauli operators conditioned on successful fusion outcomes $\psi^\pm$ are shown in Fig.~\ref{fig:3}b.
Because these operators are the stabilizers of the targeted final joint state, their measurements enable benchmarking the fusion performance by verifying the presence of entanglement 
between the early spin qubit and the late spin qubit.
In particular, they can be used to calculate the fidelity with the target state using standard analysis techniques~\cite{guhne_toolbox_2007}.
We find $\mathcal{F}=0.59(3)$ when conditioning on the fusion outcome $\psi^{+}$ and $\mathcal{F}=0.58(3)$ when conditioning on $\psi^{-}$.
Both cases are significantly above the $50\%$ bound, indicating genuine spin qubit entanglement. 
The entanglement is encoded in a single QD spin state at two different times, and interestingly the time separation is ultimately limited by the fiber propagation loss and can be much longer than the spin coherence time of the emitter. 
Indeed, between photon generations, the coherence between the spins at different times is erased by the spin initialization pulses, and the fusion operator recovers it. 
Therefore, the time-like fusion gate can be interpreted as a heralded quantum memory operation: although the spin state is collapsed and reinitialized between the two time instances, the quantum information is stored in the photon initially entangled with it, and the fusion operation effectively teleports it into the new spin state, thereby prolonging the effective coherence of the spin state.  

An important metric to benchmark the functionality of fusion operations in a network is the noise rate of the fusion outcomes, i.e., the rate with which the erroneous results are obtained in the parity checks~\cite{Bartolucci2023}. 
Due to the correspondence between the fusion outcome and the joint spin state described above, the error rates in a Pauli operator associated with a fusion outcome can be probed by analyzing the error rates in the same operator but on the spin qubits.
Note, however, that estimating fusion error rates through the spin states introduces additional imperfections due to noises (e.g. rotation and readout errors) in the spin system. 
The obtained error rates should therefore be considered upper bounds on the intrinsic performance of the photonic fusion gate.
Expectation values for the $ZZ$, $XX$, and $YY$ spin operators conditioned on the successful fusion outcomes $\psi^{\pm}$ are reported in Fig.~\ref{fig:3}c. 
For the $\psi^+$ fusions, the corresponding error rates are 18(3)\%, 31(2)\%, and 33(2)\% for $ZZ$, $XX$, and $YY$, respectively, and conditioning on $\psi^-$ leads to 19(2)\%, 33(2)\%, 33(2)\%. 
In the same figure, we also report the $ZZ$ expectation value conditioned on the fusion failure outcome $\phi^{\pm}$, obtained from detection patterns with both photons detected in either the early or the late time bin. The $XX$ and $YY$ operators are erased in this failure case and thus not reported.
These events have a higher contribution from residual background photons (see Supplementary Information~\cite{SI}), resulting in a higher $ZZ$ error rate of 32(4)\%.
In the Supplementary Information~\cite{SI} we report an analysis of physical mechanisms that contribute to the error rates and show that a large portion of the noise budget is expected to arise from spin noise. 
Routes to further improvement of the experimental performance have been discussed in detail in Ref. \cite{meng2023deterministic}.

\ \\
\noindent\textbf{Conclusion and outlook.}
We have presented photonic fusion between entangled resource states generated by quantum emitters and benchmarked the system performance through the quantum correlations of the spin states.
Our proof-of-concept demonstration may be further improved by advancing the spin system and in particular strain-free GaAs droplet-epitaxy QDs~\cite{nguyen_enhanced_2023,zaporski2023ideal} with reduced high-frequency nuclear spin noise appears an attractive route to significantly improve the system performance.
These methods promise to enable fusion-based networks with noise rates approaching the thresholds for current fault-tolerant photonic quantum computing architectures~\cite{Tiurev2022}. 
Furthermore, the optimization of tailored architectures that can take advantage of the spin-photon building block for quantum photonic hardware may bring hardware requirements closer to near-term technology~\cite{lobl2023, deGliniasty2023}.
A main advantage of the photonic approach is that large entangled states can be built from few hardware components (tens to hundreds of quantum emitters~\cite{bombin2021interleaving}) by repetitions of just two primitives: near-deterministic resource state generation and fusion operations.  
We have reported the first experimental demonstration of both functionalities constituting an important step towards scalable fusion-based photonic quantum technologies.  
%


\ \\
\noindent\textit{Acknoweldgements.}
We are grateful to Klaus Mølmer, Matthias Löbl, Love A. Petterson, and Kasper H. Nielsen for fruitful discussions. 
We acknowledge funding from the Danish National Research Foundation (Center of Excellence “Hy-Q,” grant number DNRF139), the Novo Nordisk Foundation (Challenge project "Solid-Q"), the European Union’s Horizon 2020 research and innovation program under Grant Agreement No. 820445 (project name Quantum Internet Alliance). Y.M. acknowledges funding from the European Union’s Horizon 2021 under grant agreement No. 101060143 (project name ODeLiCs). M.L.C. acknowledges funding from the European Union’s Horizon 2020 Research and Innovation programme under grant agreement No. 861097 (project name QUDOT-TECH). S.P. acknowledges financial support from the European Union’s Horizon 2020 Marie Skłodowska-Curie grant No. 101063763, from the Villum Fonden research grant No. VIL50326, and from the NNF Quantum Computing Programme.

\ \\

\noindent\textit{Conflicts of Interest.}
P.L. is a founder of the company Sparrow Quantum which commercializes single-photon sources. The authors declare no other conflicts of interest.

\noindent\textit{Data Availability.}
Data underlying the results presented in this paper are available in Ref.~\cite{online_data}.

\noindent\textit{Supplemental Document.}
See Supplementary Information ~\cite{SI} for supporting content. 
\bibliography{bib.bib}

\end{document}